 \documentclass[final,5p,times,twocolumn,authoryear]{elsarticle}

\usepackage{graphicx}
\usepackage{amssymb}
\usepackage{amsmath}
\usepackage{multirow,array}

\usepackage{lineno,hyperref}
\usepackage{color}
\hypersetup{colorlinks=true,linkcolor=red,citecolor=cyan}

\journal{Journal}

\begin{document}

\begin{frontmatter}



\title{Analyzing geodesic motion in Bocharova-Bronnikov-Melnikov-Bekenstein spacetime}


\author[1,2,3]{Bobur Turimov}\ead{bturimov@astrin.uz}
\author[1]{Ozodbek Rahimov}
\author[1]{Akbak Davlataliev}
\author[4]{Pulat Tadjimuratov}
\author[5]{Mukhiddin Norkobilov}
\author[7]{Shakhzoda Rahimova}

\affiliation[1]{Ulugh Beg Astronomical Institute, Astronomy St. 33, Tashkent 100052, Uzbekistan}
\affiliation[2]{University of Tashkent for Applied Sciences, Str. Gavhar 1, Tashkent 100149, Uzbekistan}
\affiliation[3]{Shahrisabz State Pedagogical Institute, Shahrisabz Str. 10, Shahrisabz 181301, Uzbekistan}
\affiliation[4]{Central Asian University, Milliy Bog St. 264, Tashkent 111221, Uzbekistan}
\affiliation[5]{National Research University TIIAME, Kori Niyoziy 39, Tashkent 100000, Uzbekistan}
\affiliation[6]{Joint Belarusian-Uzbek Intersectoral Institute of Applied Technical Qualifications, Kibray, Karamurt 1, Tashkent region 111200, Uzbeksitan}

\begin{abstract}
In this paper, we explored novel feature of the Bocharova-Bronnikov-Melnikov-Bekenstein (BBMB) black hole by analyzing geodesic motion. We first examined its thermodynamics and showed that Hawking temperature equals to zero. We investigated  motion of both massive and massless particles around the BBMB black hole and studied the characteristic radii, namely, marginally stable circular orbit (MSCO) and marginally bound orbit (MBO) for massive particles orbiting the BBMB black hole. Additionally, we found that the energy efficiency of massive particles in the BBMB spacetime can reach up to $8\%$. We also studied the capture cross section of massless (photon) and massive particles by the BBMB black hole. From the equations of motion, we derived the radial function crucial for determining the critical value of the impact parameter for photons and particles. Comparing these findings with the Schwarzschild spacetime, we observed significant differences in gravitational properties. Specifically, the impact parameter for a photon is smaller in the Schwarzschild field than in the BBMB field, indicating weaker gravity around the BBMB black hole, as corroborated by the closer location of the photon sphere in the BBMB spacetime. We derived explicit expressions for the pericentric precession and the deflection angle of light by the BBMB black hole, along with the trajectory of massive particles orbiting the black hole. We showed that test particles on elliptical trajectories experience pericenter shifts, with pericentric precession in the BBMB spacetime being slightly less than that predicted by Einstein's general theory of relativity. Lastly, we studied the deflection of light rays and gravitational lensing effects by the BBMB black hole in both strong and weak field approximations, incorporating general relativistic effects from the Schwarzschild spacetime. We derived expressions for the deflection angle in first and second order approximations, and used the gravitational lensing equation to determine the magnification of primary and secondary images. 

\end{abstract}



\end{frontmatter}



\section{Introduction}The BBMB black hole is a solution to the Einstein field equations coupled with a conformally coupled scalar field. This black hole solution is notable in theoretical physics for extending the standard black hole solutions of general relativity by including scalar fields, which are hypothesized in various theories beyond the Standard Model of particle physics \cite{Bocharova1970MVSFA,Bekenstein1974AP}. The extension of the BBMB solution can be found in Refs. \cite{Bekenstein1975AP,Martinez2003PRD}. The following key features characterize the BBMB black hole: (i) The solution arises from the Einstein field equations with a scalar field that is conformally coupled to gravity. This means the scalar field directly affects the curvature of spacetime. (ii) The metric of the BBMB black hole is similar to the Reissner-Nordström solution but includes additional terms related to the scalar field. The scalar field diverges on the event horizon, which presents unique challenges and insights into black hole physics. (iii) The BBMB black hole is often discussed in the context of extremal solutions, where the charge and mass parameters are finely balanced. This extremal nature influences the thermodynamic properties and stability of the black hole.

The rotating Bocharova-Bronnikov-Melnikov-Bekenstein black hole solution and also its angular as well as mass multipolar generalizations has been discussed in \cite{Astorino2015PRD}. The Israel-type proof of the uniqueness theorem for static spacetimes outside the photon surface in the Einstein-conformal scalar system has been studied in \cite{Shinohara2021PTEP}. In Ref. \cite{Senjaya2024EPJC}, an exact solution to the Klein-Gordon equation for both massive and massless scalar fields in the BBMB spacetime is provided. Furthermore, the study also discusses and calculates the Hawking radiation emanating from the horizon of the BBMB black hole using the Damour–Ruffini method. A massive particle motion in alternative theory of gravity has been investigated in \cite{Turimov2023PDU,Turimov2023PLB}.

Gravitational capture is crucial for observing the region surrounding a black hole. This phenomenon is determined by the energy and angular momentum of the particle being captured \cite{2018grav.book.....M,1983bhwd.book.....S}. To investigate gravitational capture, it is necessary to reformulate the equation of motion as a homogeneous polynomial and use this equation to estimate the impact parameter. The impact parameter is defined as the ratio of angular momentum to particle energy. Previous studies have explored similar topics in various space-time frameworks. For example, gravitational capture by a charged black hole in the Reissner-Nordström metric was examined in \cite{Zakharov1994CQG}, the capture of magnetized particles in the Schwarzschild metric was discussed in \cite{AAA:PS:2014,2021Univ....7..307A,2023Galax..11...70T,Davlataliev:2024ekv,Davlataliev:2024wdd,Rayimbaev:2020gqj,Narzilloev:2020jqn,Rayimbaev:2021luv,Abdujabbarov:2020hdp,Rayimbaev:2020izu,Rayimbaev:2014lta,Narzilloev:2021bst}, and the capture in higher-dimensional space-times within the Schwarzschild - Tangherlini metric was analyzed in \cite{2024PDU....4401483R}. This section concludes with a comparison of the effective cross-section of gravitational capture with the findings of these previous studies.

Based on astronomical observations, in the early 1600s Kepler established that the orbit described by a planet in the solar system is an ellipse, with the Sun occupying one of its focus. In fact the Keplerian laws are derived within the framework of Newtonian theory. However, this theory does not explain orbital motion of some planets in solar system, in particular, motion of Mercuiry. That is why general relativity is a very successful theory of the gravitational field suggested by Einstein in 1915, whose predictions are in excellent agreement with a large number of astronomical observations and experiments performed at the scale of the Solar System. In particular, three fundamental tests of general relativity, the perihelion precession of planet Mercury~\cite{Lo2013AJP,Park2017AJ,Rayimbaev:2021geu,Narzilloev:2021yix,Rayimbaev:2023bjs}, the bending of light by the Sun~\cite{Lebach1995PRL,Titov2015jsrs}, and the radar echo delay experiment~\cite{Shapiro2004PRL,Fomalont2009ApJ} have all fully confirmed, within the range of observational/experimental errors, the predictions of Einstein’s theory of gravity. Recently, the authors of \cite{Javed2020EPJP} calculated the deflection angle of light within a plasma medium by a BBMB black hole using the Gibbons and Werner approach (Gauss-Bonnet method). The perihelion shift in alternative theories of gravity has been explored in various studies. For instance, the perihelion precession of planetary orbits in Brans-Dicke theory was analyzed in~\cite{Weinberg:1972kfs}. In modified gravity theories like f(R) gravity, perihelion precession has been examined in \cite{Schmidt2008PRD}. The analysis of perihelion precession in the context of the Randall-Sundrum model can be found in the work of Overduin et al. (2000) \cite{Elgaroy2008}. Additionally, the perihelion shift in scalar-tensor-vector gravity (MOG) has been studied in \cite{Monica2022Universe,Monica2022MNRAS,Turimov2022MNRAS}. These references provide a comprehensive overview of how perihelion precession is affected by different alternative theories of gravity.

As we mentioned before that gravitational lensing is one of the most important tests of general relativity. It has been observed in distant astrophysical sources, however these observations are poorly controlled and it is uncertain how they constrain general relativity. The most precise tests are analogous to Eddington's 1919 experiment: they measure the deflection of radiation from a distant source by the Sun. The sources that can be most precisely analyzed are distant radio sources, in particular, some quasars are very strong radio sources. An important improvement in obtaining positional high accuracies was obtained by combining radio telescopes across Earth. The technique is called very long baseline interferometry (VLBI). With this technique radio observations couple the phase information of the radio signal observed in telescopes separated over large distances. Recently, these telescopes have measured the deflection of radio waves by the Sun to extremely high precision, confirming the amount of deflection predicted by general relativity aspect to the $0.03\%$ level. ~\cite{Bisnovatyi-Kogan:2010flt,Bozza:2002zj,Bozza:2003cp,Bezdekova:2024vct,2015MNRAS.451...17R,Fomalont2009,Davlataliev:2023ckw,Ditta:2023rhr,Pahlavon:2024caj,Atamurotov:2021qds,10.1088/1572-9494/ad6853,Turimov2019IJMPD,Rayimbaev:2024abz,Jumaniyozov:2024eah,Rahmatov:2024foj}

In the present paper, we derive the expression for the perihelion shift of the planet in the framework of general relativity and conformally coupled scalar theory. The paper is organized as follows. In Sect. ~\ref{Sec:BBMB} we discuss curvature invariant and thermodynamical properties of the BBMB black hole. In Sect. \ref{Sec:Geodesic} we study geodesic motion massive particle in the BBMB geometry. In Sect.\ref{Sec:Capture}, we discusses the gravitational capture of massive and massless particles by the BBMB black hole. In the next Sect.~\ref{Sec:Perihelion}, we provide in very detailed derivation of the pericentric precession of test particle orbiting around the BBMB black hole. Sect. \ref{Sec:FF} is devoted to the oscuillatory motion of massive particle near stable circular otbit around the black hole. In Sect.~\ref{Sec:Deflection}, we discuss the deflection of light ray and gravitational lensing effect in the general relativity. Finally, in Sect.~\ref{Sec:Conclusions}, we summarize obtained results.

\section{BBMB black hole and thermodynamic  properties}\label{Sec:BBMB}

\begin{figure}
\begin{minipage}{\columnwidth}
\centering
\includegraphics[width=\hsize]{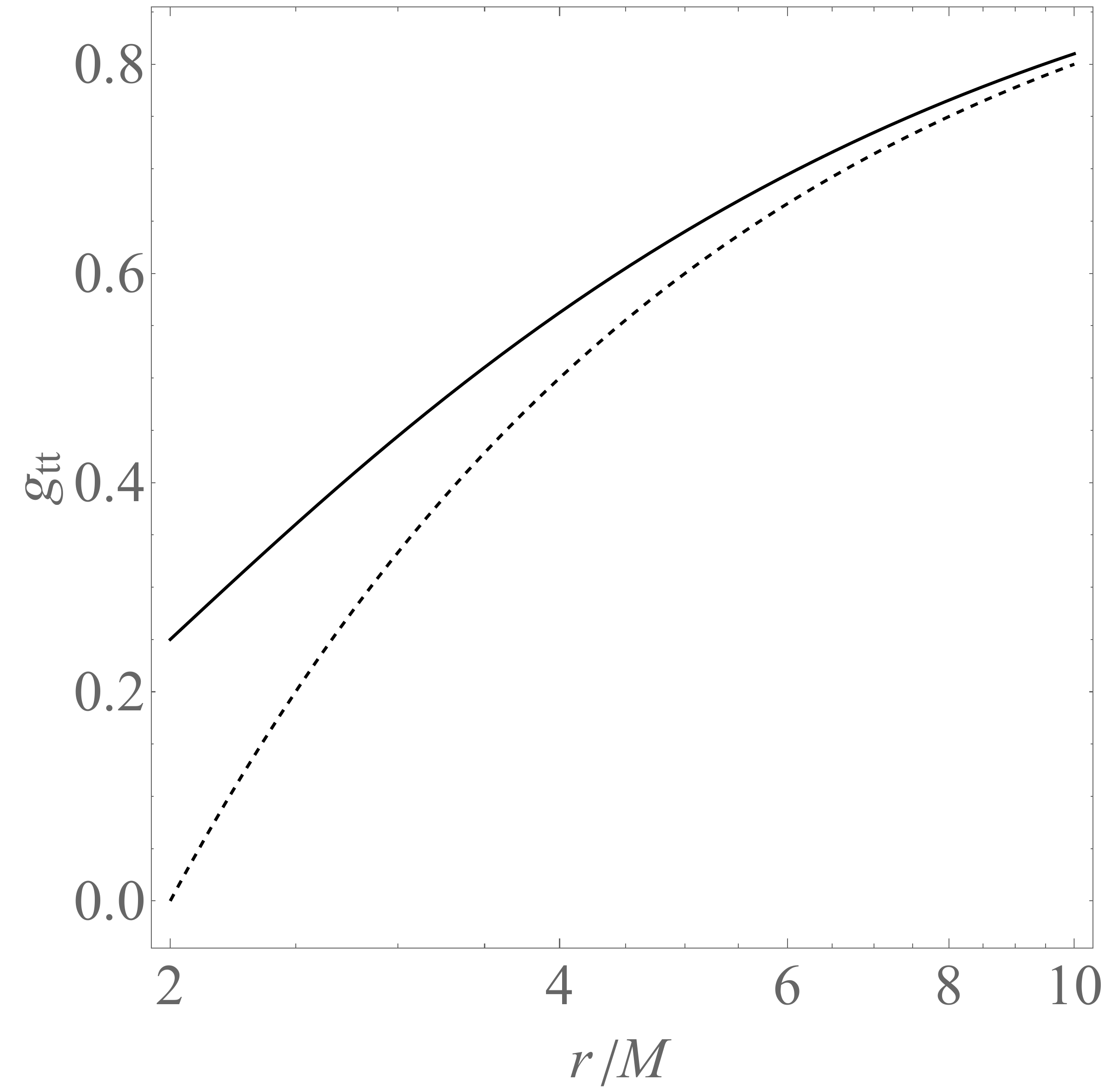}
\includegraphics[width=\hsize]{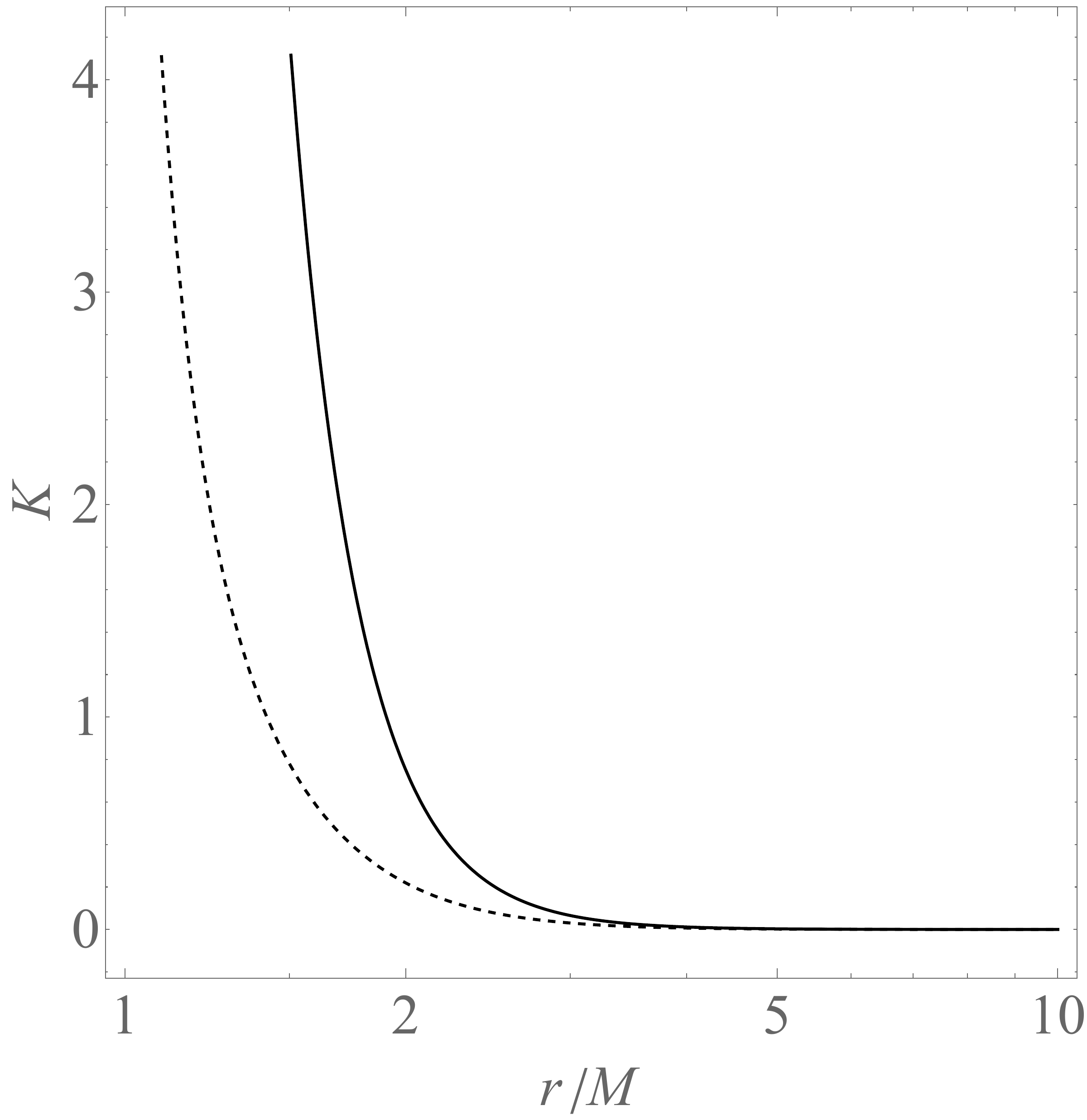}
\end{minipage}
\caption{The radial dependence of the temporal metric tensor and Kretchmann scalar. Solid line represent quantities in the BBMB spacetime while dashed line in the Schwarzschild spacetime. Blue and red points in each curves correspond to the MSCO position in the both spacetimes, respectively.\label{gtt}}
\end{figure}

The action for the Einstein conformally coupled scalar field system can be expressed as \cite{Bocharova1970MVSFA,Bekenstein1974AP}
\begin{equation}
S=\int d^4x\sqrt{-g}\left[\frac{R}{2\kappa}-\frac{1}{2}(\nabla\Phi)^2-\frac{1}{12}R\Phi^2\right]\ ,    
\end{equation}
where $\phi$ is the scalar field and $R$ is the Ricci scalar. The field equations read
\begin{align}\label{EE}
&G_{\mu\nu}=\kappa T_{\mu\nu}\ ,
\\\label{KG}
&\nabla^2\Phi=\frac{1}{6}R\Phi\ ,
\end{align}
where $G_{\mu\nu}=R_{\mu\nu}-\frac{1}{2}g_{\mu\nu}R$ is the Einstein tensor and $T_{\mu\nu}$ is the energy-momentum tensor of the system given as
\begin{align}\nonumber
T_{\mu\nu}=\frac{1}{4\pi}\left[\nabla_\mu\Phi\nabla_\nu\Phi-\frac{1}{2}g_{\mu\nu}(\nabla\Phi)^2\right.\\\left.+\frac{1}{6}\left(g_{\mu\nu}\nabla^2-\nabla_\mu\nabla_\nu+G_{\mu\nu}\right)\Phi^2\right]\ ,
\end{align}
Hereafter taking trace from equation \eqref{EE}, one obtains
\begin{align}\nonumber
-R&=\frac{\kappa}{4\pi}\left[-(\nabla\Phi)^2+\frac{1}{2}\nabla^2\Phi^2-\frac{1}{6}R\Phi^2\right]
\\
&=\frac{\kappa}{4\pi}\Phi\left[\nabla^2\Phi-\frac{1}{6}R\Phi\right]\ ,
\end{align}
and recalling equation \eqref{KG}, one can find that the Ricci scalar vanishes 
\begin{align}
R=0\ ,
\end{align}
and 
\begin{align}
\nabla^2\Phi=0\ .   
\end{align}
Hereafter re-scaling the scalar field $\Phi=\phi\sqrt{24\pi/\kappa}$, Einstein field equation can be rewritten as
\begin{align}
\left(1-\phi^2\right)R_{\mu\nu}=4\nabla_\mu\phi\nabla_\nu\phi-g_{\mu\nu}(\nabla\phi)^2-2\phi\nabla_\mu\nabla_\nu\phi ,
\end{align}
The metric of the BBMB black hole is given as \cite{Bocharova1970MVSFA,Bekenstein1974AP}
\begin{align}\label{metric}
ds^2=-\left(1-\frac{M}{r}\right)^2dt^2+\left(1-\frac{M}{r}\right)^{-2}dr^2+r^2d\Omega\ ,
\end{align}
where $M$ is the total mass of the black hole. The associated the scalar field is
\begin{align}
\phi=\frac{M}{r-M}\ .    
\end{align}
The horizon of the BBMB black hole is located at $r_h=M$, which is two times less the Schwarzschild radius, and the scalar field is divergent at the horizon. The Kretchmann scalar for given spacetime reads as follows
\begin{align}
K=\frac{48M^2}{r^6}\left(1-\frac{2M}{r}+\frac{7M^2}{r^2}\right)\ ,    
\end{align}
which contains a single singularity located at origin. At the horizon it reduces to $K(r=r_h)=288M^{-4}$. To better understand the BBMB spacetime, one can consider the radial dependence of the Kretschmann scalar and the temporal component of the metric tensor, and compare them to those in the Schwarzschild spacetime. Figure \ref{gtt} shows radial dependence of those two functions. As one see that near the black hole at fixed radial distance the Kretchman scalar in the BBMB spacetime smaller than that in the Schwarzschild spacetime. It implies that the gravitational field near the BBMB black hole in weaker than that in the Schwarzschild spacetime. 

Black hole thermodynamics is a fascinating field of study that combines principles of thermodynamics with the physics of black holes. One of the interesting feature of the black hole is Hawking temperature~\cite{Hawking1974}. Unlike the Schwarzschild black hole, one can show that Hawking temperature of the BBMB black hole is to be absolutely zero:
\begin{align}\label{Tem}
T=\frac{1}{4\pi}\frac{dg_{tt}}{dr}\Big|_{r=r_h}=0\ .  
\end{align}
It means that the BBMB black holes do not emit radiation. However, using the area law the entropy of the BBMB black hole determined as~\cite{Hawking1974} 
\begin{align}\nonumber
S=\frac{A}{4}&=\frac{1}{4}\int\sqrt{g_{\theta\theta}g_{\phi\phi}}d\theta d\phi\\&=\frac{r_h^2}{4}\int_0^{2\pi}d\phi\int_0^\pi \sin\theta d\theta=\pi M^2\ ,
\end{align}
which is four times less than that is predicted  in the Schwarzschild spacetime.  

For a black hole characterized by the three main parameters: mass $M$, electric charge $Q$, and spin parameter $J$, the first law of thermodynamics can be expressed as
\begin{align}\label{law}
dM=TdS+\Phi dQ+\Omega dJ\ ,    
\end{align}
where $\Phi$ denotes the electric potential and $\Omega$ represents the angular velocity of the black hole. In the case of a BBMB black hole, which is both static and neutral, this equation \eqref{law} is to be simplified as $dM = TdS$. Given that the temperature of the black hole is zero, as per equation \eqref{Tem}, the differential of mass becomes zero, i.e., $dM = 0$, implying that mass of the BBMB black hole $M$ is constant.

The another important properties of the black hole is its lifetime. It is contingent upon its initial mass and the mechanisms through which it can shed mass. Black holes are theoretical black holes believed to have originated in the early stages of the universe shortly after the occurrence of the Big Bang. The rate at which black hole dissipates energy can be estimated using the Stefan-Boltzmann radiation law as \cite{Turimov:2023adq} 
\begin{align}
\frac{dM}{dt}\approx\sigma AT^4\ ,
\end{align}
where $\sigma$ denotes the Stefan-Boltzmann constant. Since the temperature of the BBMB black hole is absolutely zero, one can conclude that time variation of the black hole mass is to be zero therefore $M={\rm const}$ which coincides with the first law thermodynamics. Hence, all other thermodynamics quantities of the BBMB black hole which proportional to temperature vanish.

However the free energy of the BBMB black hole doe not vanish. It is  is a thermodynamic property  that indicates the black hole's capacity to perform work. It is calculated as follows: $F = E - TS$, where $F$ represents the free energy, $E$ stands for the total energy or black hole mass $E=M$. In the realm of black holes, the free energy is closely tied to the thermodynamic stability of the black hole. A black hole is considered to be thermodynamically stable if its free energy is negative, suggesting that it can release energy and transition to a lower energy state. Conversely, if the free energy is positive, the black hole is deemed thermodynamically unstable, as it can absorb energy and expand in size. The free energy of a black hole can be determined using the Bekenstein-Hawking formula for entropy and the formula for the temperature of a black hole. By utilizing equation (\ref{Tem}), one can show that the free energy of the BBMB black hole is equal to it's total energy and by applying the definition of total energy, one can obtain the following expression for the free energy $F = M$ in the context of the black hole thermodynamics. In Ref. \cite{Senjaya2024EPJC}, the thermodynamic properties of the BBMB black hole are studied. It is demonstrated that the temperature of the black hole is zero, and as a result it is shown that no Hawking radiation is emitted from the BBMB black hole.

\section{Geodesic motion}\label{Sec:Geodesic}

The motion of a test body around a black hole is governed by the geodesic equation:
\begin{align}\label{EOM}
{\ddot x}^\alpha+\Gamma^\alpha_{\mu\nu}{\dot x}^\mu{\dot x}^\nu=0\ ,
\end{align}
where prime denotes derivative with respect to an affine parameter, and $\Gamma^\alpha_{\mu\nu}$ are the Christoffel symbols. In the background of the BBMB spacetime \eqref{metric} geodesic equations can be explicitly written as follows:
\begin{align}\label{eqt}
&{\ddot t}+\frac{2M}{r^2}\left(1-\frac{M}{r}\right)^{-1}{\dot r}{\dot t}=0\ ,
\\\nonumber\label{eqr}
&{\ddot r}-\frac{M}{r^2}\left(1-\frac{M}{r}\right)^{-1}{\dot r}^2+\frac{M}{r^2}\left(1-\frac{M}{r}\right)^3{\dot t}^2\\&-r\left(1-\frac{M}{r}\right)^2\left({\dot\theta}^2+\sin^2\theta{\dot\phi}^2\right)=0\ ,
\\\label{eqq}
&{\ddot\theta}+\frac{2}{r}{\dot r}{\dot\theta}-\sin\theta\cos\theta{\dot\phi}^2=0\ ,
\\\label{eqf}
&{\ddot\phi}+\frac{2}{r}{\dot r}{\dot\phi}+2\cot\theta{\dot\theta}{\dot\phi}=0\ .
\end{align}
For simplicity, one can consider circular motion of test particle in the equatorial plane, i.e. $\theta=\pi/2$ and $\dot\theta=0$. After integrating equations (\ref{eqt})-(\ref{eqf}), one can get
\begin{align}\label{EOM1}
&{\dot t}=\frac{E}{m}\left(1-\frac{M}{r}\right)^{-2} ,
\\
\label{EOM2}
&{\dot\phi}=\frac{L}{mr^2} ,
\\
\label{EOM3}
&m^2{\dot r}^2=E^2-\left(1-\frac{M}{r}\right)^2\left(m^2+\frac{L^2}{r^2}\right).
\end{align}
where $E$ and $L$ are constants of integration regarded to the energy and angular momentum of test particle of mass $m$. By using the conditions ${\dot r}={\ddot r}=0$, the critical value of energy and angular momentum of particle can be derived as
\begin{align}\label{EnL}
\frac{E^2}{m^2}=\frac{(r-M)^3}{r^2(r-2M)}\ , \qquad \frac{L^2}{m^2}=\frac{Mr^2}{r-2M}\ .
\end{align}
and the stationary point of these quantities are located at a position of the marginally stable circular orbit (MSCO) for a test particle which is equal to $ r_{\rm ms}=4M$, while in the Schwarzschild spacetime it equals to $r_{\rm ms}=6M$. The MSCO, also known as the innermost stable circular orbit (ISCO), is a critical concept in the study of accretion disks around compact objects like black holes and neutron stars. The MSCO represents the smallest orbit in which a test particle can stably circle a massive object without eventually spiraling inward due to the object's gravitational influence. The MSCO marks the transition between stable and unstable orbits. Inside this radius, any perturbation can lead to the particle quickly plunging into the black hole. For accretion disks, the MSCO plays a crucial role in determining the inner edge of the disk and, consequently, the dynamics and emission characteristics of the disk. The total energy of massive particle at the MSCO position around the BBMB black hole is determined as 
\begin{align}
E_{\rm ms}=\frac{3\sqrt{6}}{8}m\ ,
\end{align}
which is less that rest energy. Similarly, the angular momentum of particle at the same orbit is $L_{\rm ms}=4mM$.
\begin{figure}
\begin{minipage}{\columnwidth}
\centering
\includegraphics[width=\hsize]{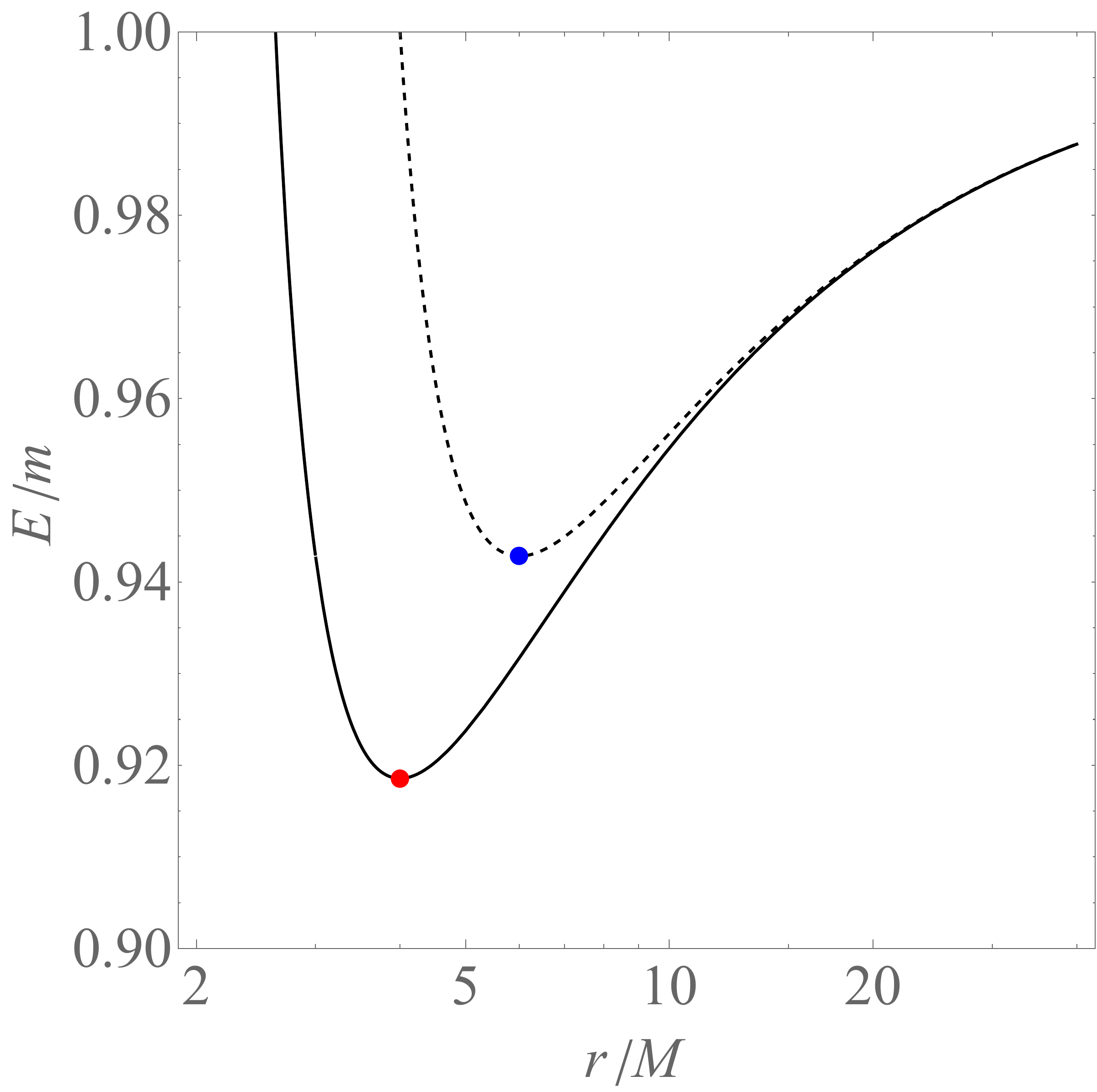}
\includegraphics[width=\hsize]{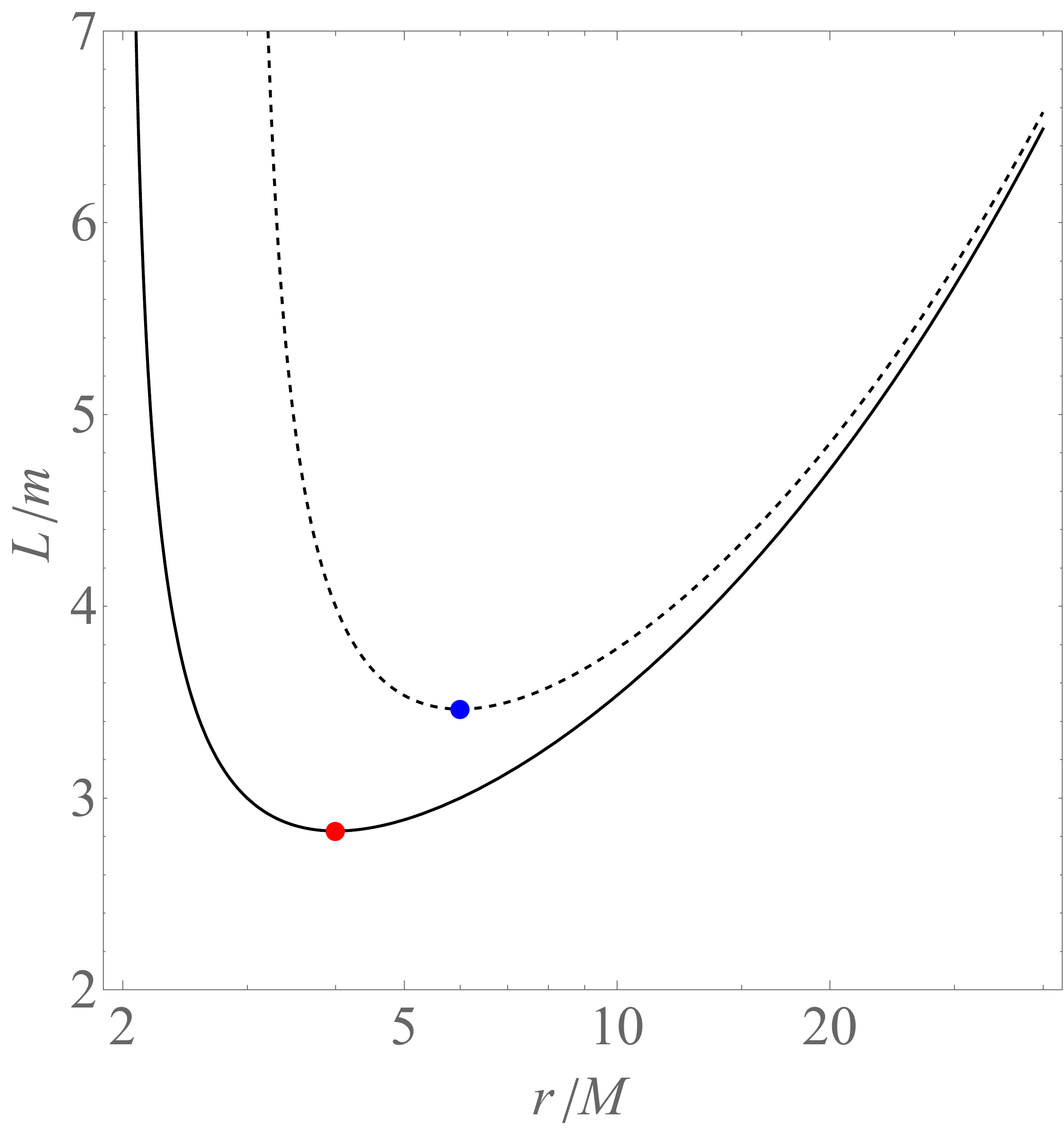}
\end{minipage}
\caption{The radial dependence of the critical energy and angular momentum of particle. Solid line represent quantities in the BBMB spacetime while dashed line in the Schwarzschild spacetime. Blue and red points in each curves correspond to the MSCO position in the both spacetimes, respectively. \label{EL}}
\end{figure}

A marginally bound circular orbit (MBO), is a specific type of orbit in the context of general relativity and the motion of particles around compact objects like black holes. This orbit represents the critical boundary between bound and unbound trajectories. Here are some key points about MBO: A marginally bound orbit is the orbit where a particle has just enough energy to remain bound to the black hole or neutron star. Any slight increase in energy would cause the particle to escape to infinity, while any slight decrease would cause it to spiral into the compact object. In the MBO, the total energy of the particle  equals the rest energy. It can be determined from following condition $E=m$ in equation \eqref{EnL}. In particular, in the BBMB spacetime the MBO radius can be determined as 
\begin{align} 
r_{\rm mb}=\frac{1}{2}(3+\sqrt{5})M\ .
\end{align}
which is less than $4M$ predicted in Schwarzschild spacetime.

Energy efficiency, also known as gravitational defect mass, is a critical concept in the dynamics of particles in strong gravitational fields. It provides insights into how much energy can be extracted or lost due to the gravitational influence of a compact object, such as a black hole or neutron star. High energy efficiency implies that a significant amount of the particle's energy can be extracted or lost in the process of approaching or interacting with the MSCO. This is relevant for understanding energy conversion mechanisms in accretion disks around black holes and neutron stars. The energy efficiency is crucial for modeling phenomena such as high-energy radiation from accretion disks, relativistic jets, and the emission spectra of compact objects. The concept of gravitational defect mass is related to how much of the particle’s energy is effectively "defected" or altered by gravitational effects. This term helps quantify the impact of gravitational fields on particle energy. The energy efficiency of particle in the strong gravitational field is found as $\eta=1-{\cal E}_{MSCO}$, where ${\cal E}_{\rm MSCO}$ is the specific energy of particle at the MSCO. In the BBMB spacetime, it can be determined as 
\begin{align}
\eta=1-\frac{3\sqrt{6}}{8}\simeq 0.08\ ,    
\end{align}
and in the Schwarzschild spacetime the energ efficiency is approximately $\eta\simeq 0.06$.

\section{Capture cross section by BBMB black hole}\label{Sec:Capture}

The capture cross section of a black hole is a fundamental concept in astrophysics that describes the effective area through which particles or light rays can be captured by the black hole's gravitational field \cite{Zakharov1994CQG,2021Galax...9...65T}. This measure is crucial for understanding various astrophysical phenomena, including accretion processes, gravitational lensing, and the formation of black hole shadows. The capture cross-section can be thought of as the "target area" that a black hole presents to incoming particles or photons. If a particle or photon enters this area, it will be unable to escape the gravitational field of the black hole and will eventually be captured by it. Mathematically, the capture cross-section is related to the impact parameter $b$, which is the perpendicular distance from the center of the black hole to the trajectory of an incoming particle or photon (i.e. $\sigma=\pi b^2$). Here we consider this problem in the BBMB spacetime. From equation \eqref{EOM3} equation for radial motion reduces to
\begin{align}\nonumber\label{R}
m^2r^4{\dot r}^2&=\left(E^2-m^2\right)r^4+2m^2Mr^3\\&-\left(L^2+m^2M^2\right)r^2+2L^2Mr-L^2M^2=R(r)\ .   
\end{align}

\subsection{Photon's case} 

\begin{figure}
\begin{minipage}{\columnwidth}
\centering
\includegraphics[width=\hsize]{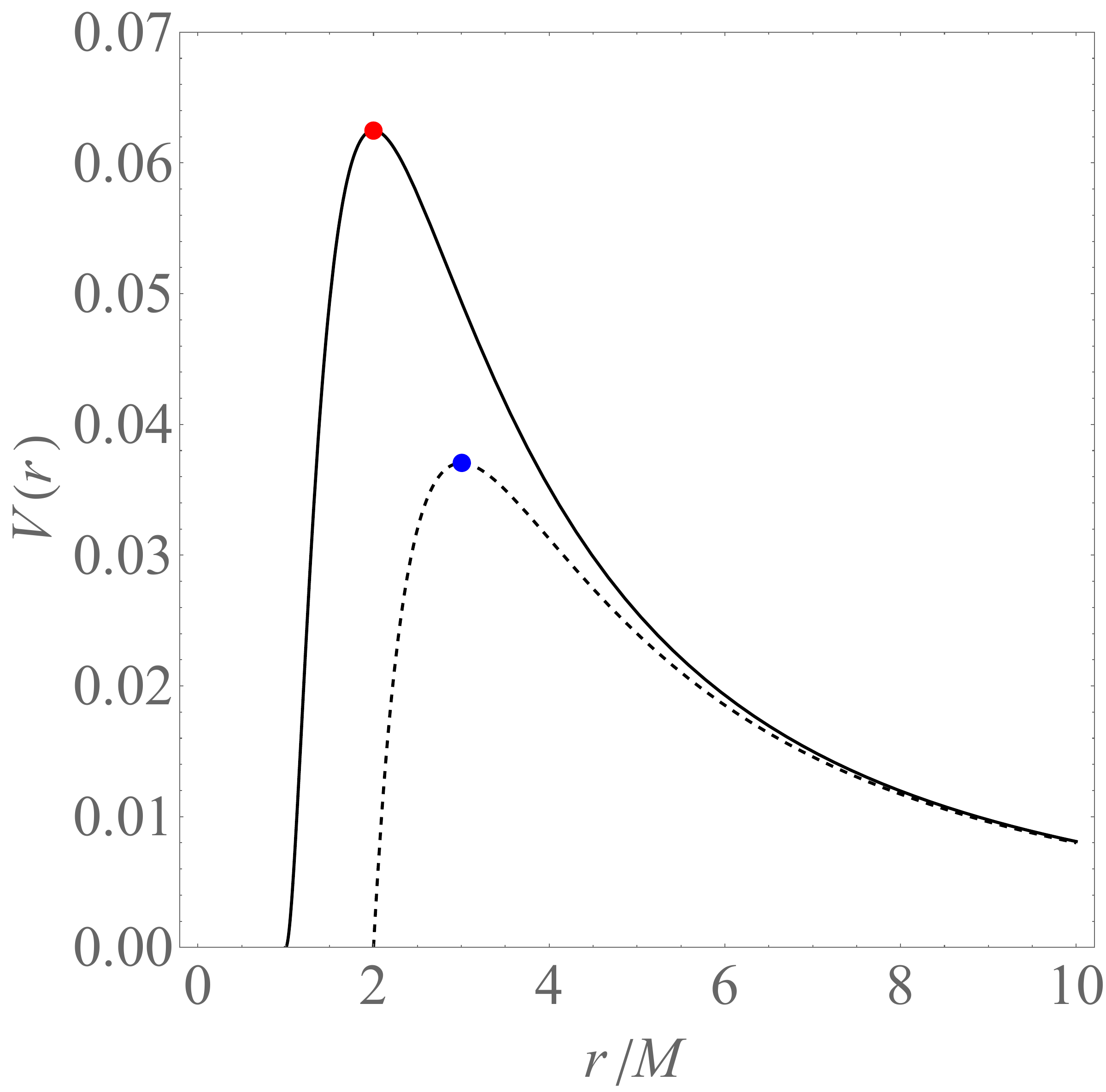}
\end{minipage}
\caption{The radial dependence of the effective potential for photon in the BBMB (solid line) and Schwarzschild (dashed line) spacetimes. \label{effpot}}
\end{figure}

The cross-section of a photon being absorbed or scattered by a black hole is an important task in astrophysics. In the case of photon (i.e. $m=0$) radial function in equation \eqref{R} can be written as follows

\begin{align}\nonumber\label{captureq}
\frac{R(r)}{E^2}&=r^4-b^2(r-M)^2\\&=[r^2+b(r-M)][r^2-b(r-M)]\ . 
\end{align}
where $b=L/E$ is the impact factor of photon. From the condition following $R(r)=0$, the following quadratic equation can obtained as
\begin{align}
r^2\pm br\mp bM=0\ ,\quad\to\quad r=\frac{\pm b\pm\sqrt{b^2\mp 4bM}}{2} \ ,   
\end{align}
and the critical value of the radial coordinates can be determined by setting the expression inside the square root to zero. From this fact critical impact parameter of photon is determined as $b=4M$ and radius of the photonsphere is given as $r_{\rm ph}=2M$. On the other hand photon motion can be descibed the following equation 
\begin{align}
\left(\frac{dr}{d\lambda}\right)^2=\frac{1}{b^2}-V(r)\ ,    
\end{align}
where $\lambda$ is an affine parameter and $V(r)$ is the effective potential for photon defined as
\begin{align}\label{V}
V(r)=\frac{1}{r^2}\left(1-\frac{M}{r}\right)^2\ .    
\end{align}
After performing simple algebraic calculations, one can easily find that the maximum value of the effective potential for a photon in the BBMB spacetime is $ V_{\rm max}=b^{-2}=1/16M^2 $ at $ b=4M $. In the Schwarzschild spacetime, the effective potential for photon has a maximum at $V_{\rm max}=1/27M^2$ with an impact parameter of $b=3\sqrt{3}M$. In order to compare our result that is presented in the Schwarzschild spacetime the radial dependence of the effective potential in equation \eqref{V} is illustrated in Fig.\ref{effpot}. From these results, it is evident that the stationary point of the effective potential corresponds to the position of the photon sphere in both the BBMB and Schwarzschild spacetimes. Additionally, the maximum of the effective potential represents the inverse square of the photon's impact parameter.

\begin{figure}
    \centering
    \includegraphics[width=\hsize]{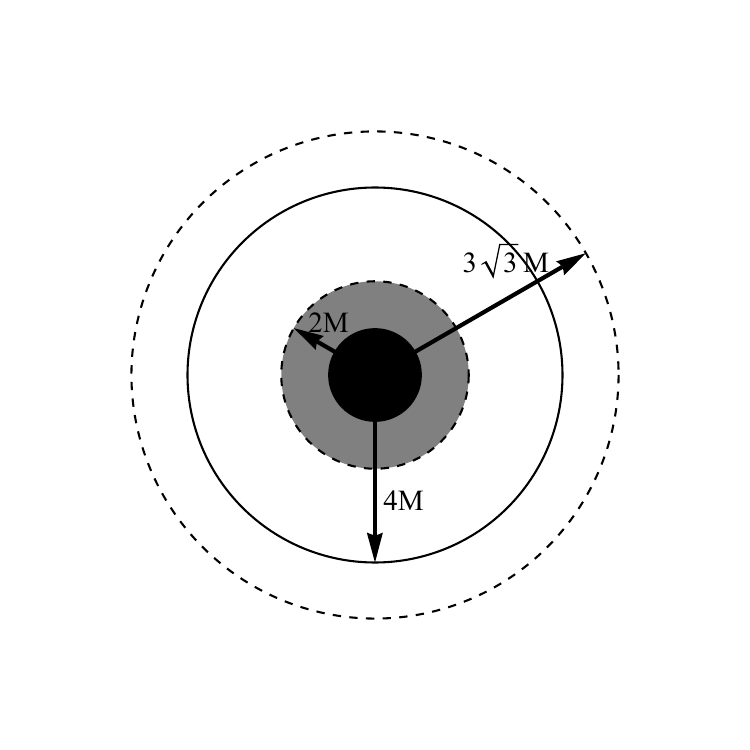}
    \caption{A comparison of the capture cross section for photon in both BBMB (solid line) and Schwarzschild (dashed line) spacetimes, respectively. A black shaved area represents horizon of the BBMB black hole while shaved gray area represents the horizon of the Schwarzschild spacetime.}
    \label{CapturePhoton}
\end{figure}

All of our analytical and graphical analyses clearly show that the maximum value of the impact parameter or effective potential differs from the classical value (Schwarzschild case). This indicates that the gravitational properties of the BBMB black hole are significantly different from those of the Schwarzschild black hole. Specifically, the impact parameter $ b $ is $ 3\sqrt{3}M $ in the case of the Schwarzschild black hole and $ 4M $ in the case of the BBMB black hole. From this, it can be seen that the impact parameter for a photon decreases more in the field of a Schwarzschild black hole than in the BBMB case. This, in turn, means that gravity around the BBMB black hole is weaker than around the Schwarzschild black hole. This can be verified by the fact that the photon sphere around the BBMB black hole exists at a distance of $ r_{\rm ph}=2M $, whereas in the Schwarzschild case, it is at a distance of $ r_{\rm ph}=3M$. Finally, capture cross sections of photon by the BBMB and Schwarzschild black hole are 
\begin{align}
\sigma_{\rm BBMB}=16\pi M^2\ ,\qquad  \sigma_{\rm Schw.}=27\pi M^2\ .   
\end{align}
In Fig. \ref{CapturePhoton}, we show the capture cross section of photons by both the BBMB black hole and the Schwarzschild black hole. It can be observed that, due to the weaker gravitational field of the BBMB black hole compared to the Schwarzschild black hole, the size of the capture cross section is smaller in the BBMB spacetime than in the Schwarzschild spacetime.

\subsection{Slow particle's case} 

\begin{figure}
    \centering
    \includegraphics[width=\hsize]{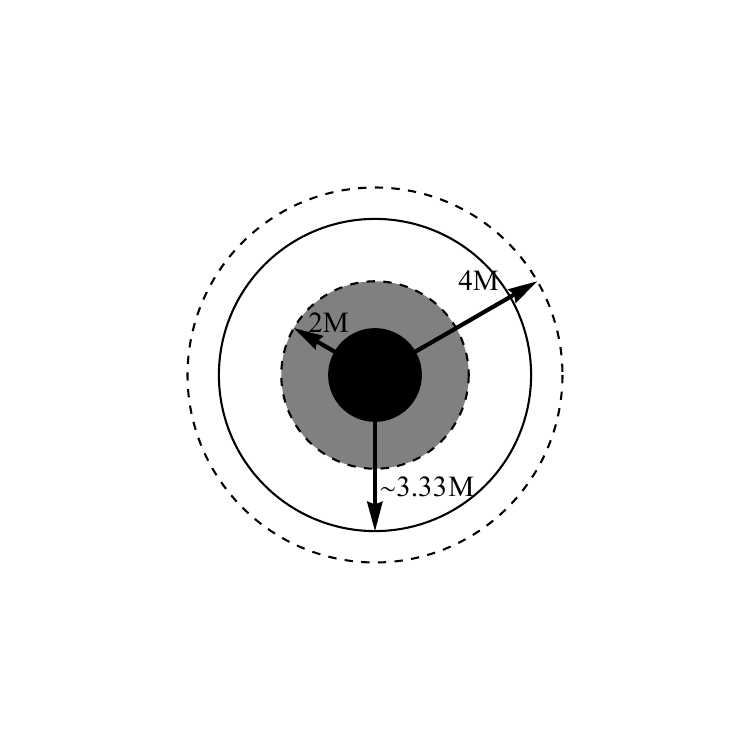}
    \caption{A comparison of the capture cross section for slow particle in the BBMB (solid line) spacetime with a raius of $\sqrt{(11+5\sqrt{5})/2}M\simeq 3.33M$ and in the Schwarzschild (dashed line) spacetime with radius of $4M$, respectively. A black shaved area represents horizon of the BBMB black hole while shaved gray area represents the horizon of the Schwarzschild spacetime.}
    \label{CaptureSlow}
\end{figure}

Now we consider capture cross section of a slow particle by the BBMB black hole. In this case, the total energy of particle will be the same as rest energy of particle, i.e. $E\simeq m$. The same considerations were made in Zakharov's work for particles capture by the Reissner-Nordstr\"om black hole \cite{Zakharov1994CQG}. Finally, the main equation (\ref{R}) can be rewrite as follows
\begin{align}\label{slow}
\frac{R(r)}{2Mm^2}=r^3-\frac{ \left(L^2+M^2m^2\right)}{2M}r^2+\frac{L^2}{m^2}r-\frac{L^2M}{2m^2}=0\ .
\end{align}
This equation characterizes a particle with angular momentum $L$. We are particularly interested in the critical value of the angular momentum of a particle when it moves around a black hole. 
The properties of the roots of a homogeneous polynomial with its coefficients are interconnected algebraically. And in the same way, equation (\ref{slow}) can be solved. One such method is given by the Sylvester matrix (See e.g.\cite{Akritas2014,2021Univ....7..307A}), which allows one to calculate the resultant of two polynomials. In this case, the second polynomial is the derivative of equation (\ref{slow}). According to the Sylvester matrix, we must find a solution to the determinant five by five.
Based on these considerations, we will simplify the solutions to equation \eqref{slow} and it has a real solution when the angular momentum satisfies the following equation:
\begin{align}\label{deteq}
L^6-11 L^4 m^2 M^2-L^2 m^4 M^4=0\ .
\end{align}
The real solutions of equation (\ref{deteq}) are regarded as the impact parameter of slow particle:  
\begin{align}\label{detangl}
b=\frac{L}{m}=\sqrt{\frac{5\sqrt{5}+11}{2}}M\ .
\end{align}
The same result can be obtained in the case of the Reissner-Nordstrom black hole, for an extremely charged state (See e.g.\cite{Zakharov1994CQG}). The capture cross section of slow particle in the BBMB spacetime is
\begin{align}
\sigma=\left(\frac{5\sqrt{5}+11}{2}\right)\pi M^2\ ,    
\end{align}
and in the Schwarzschild spacetime this quantity equals to $\sigma=16\pi M^2$. Finally, in Fig. \ref{CaptureSlow}, the capture cross section of a slow particle is illustrated for both BBMB and Schwarzschild spacetimes. The same scenario observed in the photon case can be seen here as well. 

\subsection{Massive particle case} 

\begin{figure}
    \centering
    \includegraphics[width=\hsize]{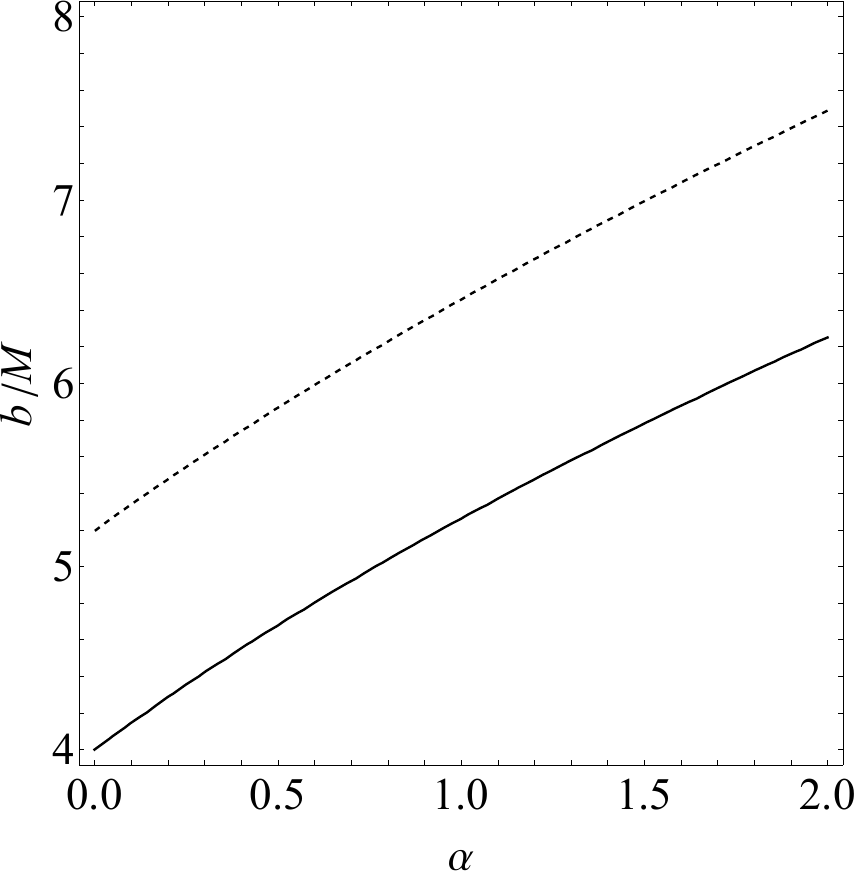}
    \caption{A comparison of the capture cross section for photon in both BBMB (solid line) and Schwarzschild (dashed line) spacetimes, respectively. A black shaved area represents horizon of the BBMB black hole while shaved gray area represents the horizon of the Schwarzschild spacetime.}
    \label{impact}
\end{figure}

To consider capture cross section of massive particle by the BBMB spacetime, one has to determine the impact parameter as follows: $b^2=L^2/(E^2-m^2)$ (See. e.g. \cite{Zakharov1994CQG}). The radial function reduces to
\begin{align}\label{captureq11}
\frac{R(r)}{E^2-m^2}&=r^4+2\alpha Mr^3-\alpha M^2r^2-b^2(r-M)^2\ , \end{align}
where $\alpha=m^2/(E^2-m^2)$ is a dimensionless parameter. The critical value of the impact parameter is determined from the following cubic equation:
\begin{align}\label{l}
l^3+\left(\alpha^2-12\alpha-16\right)l^2-\alpha^2(11\alpha +8)l-\alpha^4=0\ ,   
\end{align}
where $l=(b/M)^2$. It is known that any cubic equation has either three real solutions or one real and two imaginary solutions. Here, we do not present the explicit form of the solution to equation \eqref{l}. However, using a numerical method, the dependence of the impact parameter of a massive particle on the $\alpha$ parameter is shown in Fig. \ref{impact}. As can be seen from the results, the impact parameter of the massive particle increases due to the effect of the $\alpha$ parameter. In the absence of this parameter, the impact parameter of the massive particle is equal to that of a photon, which is $b=4M$ in the BBMB spacetime and $3\sqrt{3}M$ in the Schwarzschild spacetime (at $\alpha=0$). In case of ultra relativistic particle i.e. $E\gg m$, capture cross section in the BBMB spacetime can be estimated as
\begin{align}
\sigma=16\pi M^2\left(1+\frac{3m^2}{4E^2}\right)\ ,     
\end{align}
while in the Schwarzschild spacetime, it reduces to
\begin{align}
\sigma=27\pi M^2\left(1+\frac{5m^2}{9E^2}\right)\ .     
\end{align}

\section{Pericentric precession}\label{Sec:Perihelion}

\begin{figure}
\includegraphics[width=\hsize]{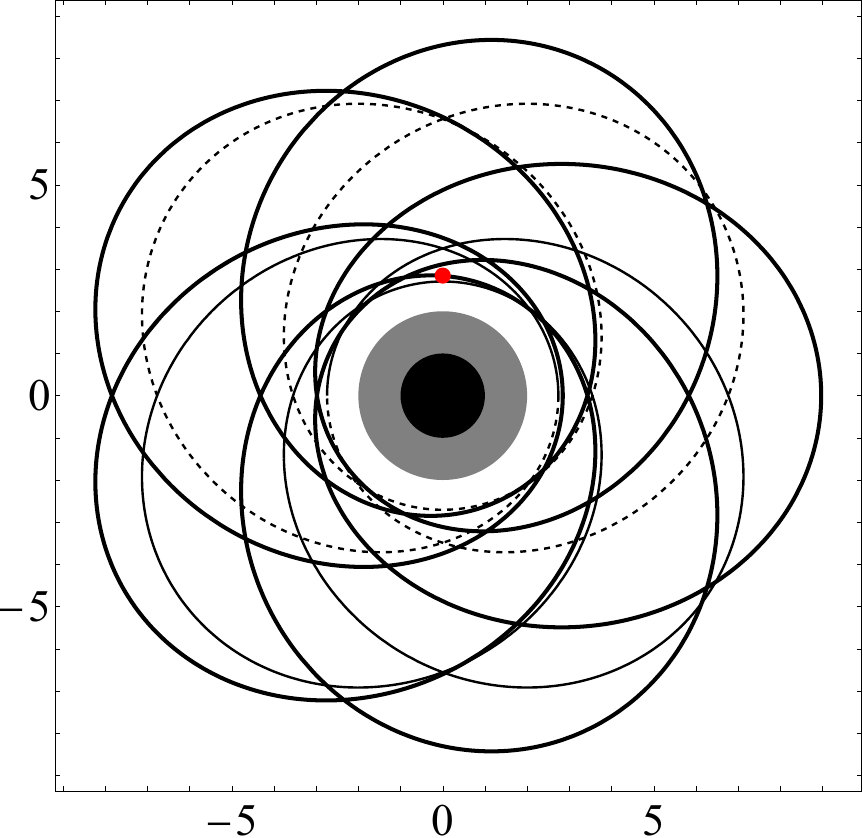}
\caption{The perihelion shift of test particle orbiting the BBMB (solid line) and Schwarzschild (dashed line) black holes for fixed values of eccentricity $e=0.6$ and expansion parameter $\epsilon=0.2$. \label{Perihelion}}	
\end{figure}

Pericentric precession, also known as apsidal or perihelion precession, refers to the gradual shift or rotation of the orbit of a celestial body around a primary mass, such as a star, planet, or black hole. This phenomenon occurs due to various gravitational influences and relativistic effects, causing the point of closest approach (pericenter) in the orbit to slowly move over time. The perihelion represents the orbit of a planet closest to the central object. This phenomenon is also one first test of the Einstein theory of relativity. It is also interesting to consider pericentric precession in the BBMB spacetime. Hereafter, introducing new dimensionless radial variable~\cite{Turimov2022MNRAS} 
\begin{align}
u= \frac{L^2}{Mm^2r}\ ,    
\end{align} 
a ratio of equations (\ref{EOM2}) and (\ref{EOM3}) reduces to
\begin{align}\label{trac}
\left(\frac{du}{d\varphi}\right)^2=\frac{E^2L^2}{M^2m^4}-\left(1-\frac{M^2m^2u}{L^2}\right)^2\left(u^2+\frac{L^2}{M^2m^2}\right) .    
\end{align}
and after differentiating from both side equation (\ref{trac}) and performing simple algebra one can get
\begin{align}\label{GR}
\frac{d^2u}{d\phi^2}+u=1-\frac{M^2m^2}{L^2}u+\frac{3M^2m^2}{L^2}u^2-\frac{2M^4m^4}{L^4}u^3\ . 
\end{align}
Here, the quadratic term with respect to $ u $ in equation (\ref{GR}) accounts for the general relativistic correction, while the linear and third-order terms arise due to the higher-order corrections in the BBMB spacetime. Mathematically, equation (\ref{GR}) represents a nonlinear harmonic oscillator equation and finding exact solution is rather difficult. Using perturbation theory, a semi-analytical approach can be applied to find the approximate value of the perihelion shift of a test particle in the BBMB spacetime. By introducing the small parameter $\epsilon=3(Mm/L)^2\ll 1$, equation \eqref{GR} is simplified as follows
\begin{align}\label{GR1}
\frac{d^2u}{d\phi^2}&+u=1-\frac{1}{3}\epsilon u+\epsilon u^2\ ,
\end{align}
while the solution can be expanded in the power of the small parameter: 
\begin{align}\label{SOL}
u(\phi)=u_0(\phi)+\epsilon u_1(\phi)+{\cal O}(\epsilon^2)\ .
\end{align}
Inserting equation (\ref{SOL}) into (\ref{GR1}), the zeroth-order approximation equation can be derived as: \[\frac{d^2 u_0}{d\phi^2} + u_0 = 1\ ,\] 
and the solution to this equation takes a form:
\[ u_0 = 1 + e \cos\phi\ , \]
which is the same result predicted by Newtonian theory, where $e$ is the eccentricity. In the first-order approximation, equation (\ref{GR1}) reads:
\begin{align}\nonumber\label{eq1}
\frac{d^2u_1}{d\phi^2}+u_1&=u_0^2-\frac{1}{3}u_0\\&=\frac{2}{3}+\frac{e^2}{2}+\frac{5}{3}e\cos\phi+\frac{e^2}{2}\cos 2\phi\ .
\end{align}
As can be observed, equation (\ref{eq1}) is a second-order non-homogeneous differential equation for $ u_1 $. The general solution to this equation is the sum of the homogeneous and particular solutions. In this case, we are not interested in the homogeneous solution for $ u_1 $ because it will be the same as that obtained in the Newtonian approximation. The particular solution can be found as:
\[u_1(\phi)=A+B\phi\sin\phi+C\cos 2\phi\ ,\]
which satisfies the following equation:
\begin{align}\label{eqcons}
\frac{d^2u_1}{d\phi^2}+u_1=A+2B\cos\phi-3C\cos 2\phi\ ,    
\end{align}
where $A$, $B$, and $C$ are unknown constants determined by comparing equations \eqref{eq1} and \eqref{eqcons}:
\begin{align}
A=\frac{2}{3}+\frac{e^2}{2}, \quad B=\frac{5e}{6}, \quad C=-\frac{e^2}{6}\ . 
\end{align}
Before presenting the final result for $u$, the following useful expression for small $\epsilon$ can be utilized:
\begin{align}\label{useful}
\cos[\phi(1-\frac{5}{6}\epsilon)]&\simeq\cos\phi+\frac{5}{6}\epsilon\phi\sin\phi+{\cal O}(\epsilon^2)\ .    
\end{align}
Consequently, considering all the aforementioned facts along with equation (\ref{useful}), the solution (\ref{SOL}) can be rewritten as:
\begin{align}
u(\phi)&=1+e\cos\left[\phi(1-\frac{5}{6}\epsilon)\right]+\frac{\epsilon}{3}\left[2+e^2\left(1+\sin^2\phi\right)\right]\ .
\end{align}
Since the precession results from the orbit not being periodic in $ 2\pi $, it must be derived from this term, and thus can be calculated accordingly. Denote the precession by $ \delta\phi $. This gives $ u(0) = u(2\pi + \delta\phi) $. The perihelion first occurs at $ \phi = 0 $, as defined earlier. This means that the second perihelion will occur when the cosine term generating the precession has gone through a full $ 2\pi $. Consequently, one can derive the following relation:
\[2\pi = (2\pi + \delta\phi)\left(1 - \frac{5\epsilon}{6}\right)\ ,\quad\to\quad \delta\phi\simeq\frac{5\pi\epsilon}{3}\ , \]
which implies that pericentric shift in the BBMB spacetime reads 
\begin{align}
\delta\phi\simeq\frac{5\pi M^2m^2}{L^2}\ .    
\end{align}
However, in the Schwarzschild spacetime it is approximately equals to 
\begin{align}
\delta\phi\simeq \frac{6\pi M^2m^2}{L^2}\ .
\end{align}
Figure~\ref{Perihelion} shows the perihelion precession of a massive particle orbiting the BBMB black hole and the Schwarzschild black hole, with fixed values for the eccentricity and expansion parameter. As expected, the pericentric precession of massive test particle in the BBMB spacetime is less than that in the Schwarzschild spacetime due to the weak gravitational field.  

\section{Orbital and epicyclic frequencies}\label{Sec:FF}

\begin{figure}
\includegraphics[width=\hsize]{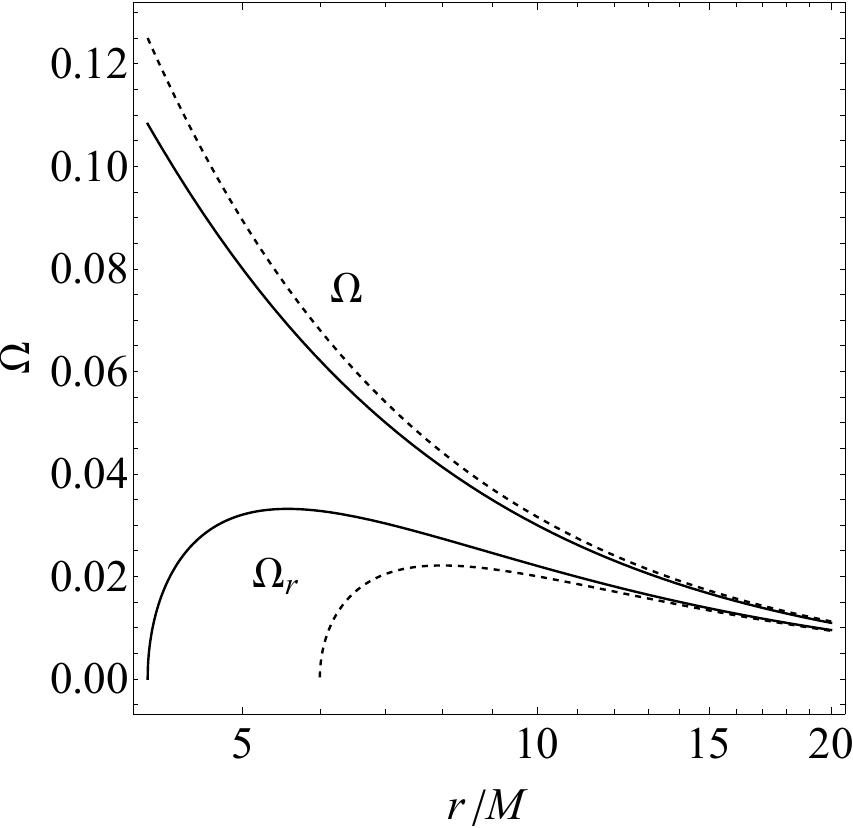}
\caption{The radil dependence of the orbital and epicyclic frequencies in the BBMB (solid line) and Schwarzschild (dashed line) spacetimes.}\label{Omega}	
\end{figure}

Orbital and epicyclic frequencies, also known as fundamental frequencies, are key concepts in the study of motion around black holes, particularly in understanding the dynamics of particles and accretion disks in strong gravitational fields. For the static spherically-symmetric spacetime (i.e., $g_{t\phi}=0$) orbital and  epicyclic frequencies can be determined as \cite{Turimov2022Universe,Turimov2020PRD}  
\begin{align}
\Omega^2&=-\frac{\partial_rg_{tt}}{\partial_rg_{\phi\phi}}\ ,
\\
\Omega_r^2&=\frac{1}{2g_{rr}}\left(g_{tt}^2\partial_r^2 g^{tt}+\Omega^2g_{\phi\phi}^2\partial_r^2 g^{\phi\phi}\right)\ ,
\\
\Omega_\theta^2&=\frac{1}{2g_{\theta\theta}}\left(g_{tt}^2\partial_\theta^2 g^{tt}+\Omega^2g_{\phi\phi}^2\partial_\theta^2 g^{\phi\phi}\right)\ ,
\end{align}
and in the BBMB spacetime they are simplified as
\begin{align}\label{W}
&\Omega=\Omega_\theta=\sqrt{\frac{M}{r^3}\left(1-\frac{M}{r}\right)}\ ,
\\\label{Wr}
&\Omega_r=\Omega\sqrt{\left(1-\frac{M}{r}\right)\left(1-\frac{4M}{r}\right)}\ ,    
\end{align}
while in the Schwarzschild spacetime these frequencies are 
\begin{align}
&\Omega=\Omega_\theta=\sqrt{\frac{M}{r^3}}\ , \qquad \Omega_r=\Omega\sqrt{1-\frac{6M}{r}}\ ,    
\end{align}
Notice that in both BBMB and Schwarzschild spacetimes the orbital and vertical frequencies equals to each other. As one can see from equation \eqref{W} that in the BBMB spacetime, the factor $\sqrt{1-M/r}$ reduces the frequency compared to the Schwarzschild case. This indicates that the gravitational influence in the BBMB spacetime is weaker, resulting in lower orbital and vertical epicyclic frequencies for the same radial distance $r$. However, in the BBMB spacetime, the radial epicyclic frequency is modified by two factors, $1-M/r$. This leads to a different behavior in radial stability. In particular, at larger distances $r \gg M$, both terms approach 1, and the frequencies become similar to the Schwarzschild case, while at smaller distances $r \sim M$, the BBMB spacetime frequencies are reduced more significantly than in the Schwarzschild case, indicating weaker radial stability near the black hole. Comparing the orbital and epicyclic frequencies in the BBMB and Schwarzschild spacetimes reveals that the gravitational properties of the BBMB black hole are notably weaker. The BBMB spacetime's additional factors reduce both the orbital and epicyclic frequencies, indicating less intense gravitational effects compared to the Schwarzschild black hole. This distinction is particularly evident in the radial epicyclic frequency, where the BBMB spacetime introduces additional weakening factors, leading to different stability characteristics for orbits near the black hole. Figure \ref{Omega} shows the radial dependence of the fundamental frequencies in both BBMB and Schwarzschild spacetime. 

It is also interesting to consider linear velocity of massive particle orbiting the BBMB black hole. Using the following definition \cite{Turimov2022Universe}:
\begin{align}
v=\sqrt{-\frac{g_{\phi\phi}}{g_{tt}}}\frac{d\phi}{dt}=\sqrt{-\frac{g_{\phi\phi}}{g_{tt}}}\Omega \ ,   
\end{align}
the linear velocity of massive particle in the BBMB spacetime can be determined as
\begin{align}
v=\sqrt{\frac{M}{r-M}} \ ,   
\end{align}
while in the Schwarzschild spacetime it is given as \cite{Turimov2022Universe}
\begin{align}
v=\sqrt{\frac{M}{r-2M}} \ ,   
\end{align}
The linear velocity of massive particle orbiting in the stable circular orbit around the BBMB and Schwzrzschild black holes are
\begin{align}
v=\frac{c}{\sqrt{3}} \ , \qquad\qquad v=\frac{c}{2}\ ,
\end{align}
where $v$ is the speed of the light. One can conclude that test particle orbits around the BBMB black hole faster than those around the Schwarzschild black hole.

\section{Gravitational lensing effect}\label{Sec:Deflection}

\subsection{Strong lensing}

Gravitational lensing, in particular, determining the deflection angle is one of the first test of the general relativity. It is also interesting to study gravitational lensing effects in the BBMB spacetime. The photon motion can be described by the following equations: 
\begin{align}
&\frac{dt}{d\lambda}=\frac{1}{b}\left(1-\frac{M}{r}\right)^{-2}\ ,
\\
&\frac{d\phi}{d\lambda}=\frac{1}{r^2}\ ,
\\
&\frac{dr}{d\lambda}=\sqrt{\frac{1}{b^2}-\frac{1}{r^2}\left(1-\frac{M}{r}\right)^2}\ .    
\end{align}
Using these expression the deflection angle and time delay can be determined as
\begin{align}
\alpha&=\pi-\int_{r_{\rm ph}}^\infty\frac{dr}{r^2\sqrt{\frac{1}{b^2}-\frac{1}{r^2}\left(1-\frac{M}{r}\right)^2}}\ ,
\end{align}
and 
\begin{align}
t=\int_{r_1}^{r_2}\frac{dr}{\left(1-\frac{M}{r}\right)^{2}\sqrt{1-\frac{b^2}{r^2}\left(1-\frac{M}{r}\right)^2}}\ .
\end{align}

Before evaluating both integrals, one can conclude that they will expressed in terms of the elliptic integral. After introducing new dimensionless variable $u=M/r$, the expression for the deflection angle reduces to 
\begin{align}\nonumber
\alpha&=\pi-\int_0^{1/2}\frac{du}{\sqrt{\frac{M^2}{b^2}-u^2\left(1-u\right)^2}}\\\nonumber&=\pi-\int_0^{1/2}\frac{du}{\sqrt{\left[\frac{M}{b}-u\left(1-u\right)\right]\left[\frac{M}{b}+u\left(1-u\right)\right]}}\\\nonumber
&=\pi+\frac{4}{\sqrt{1+\frac{M}{b}}-\sqrt{1-\frac{M}{b}}}\\&\times\left[F\left(\arcsin\eta\Bigg|\frac{1}{\eta^4}\right)+F\left(\arcsin\eta\sqrt{\frac{u_-}{u_+}}\Bigg|\frac{1}{\eta^4}\right)\right]\ ,
\end{align}
where $F(x|a)$ is the incomplete elliptic integral of the first kind and
\begin{align}\nonumber
&u_\pm=\frac{1}{2}\left(1\pm\sqrt{1-\frac{4M}{b}}\right) ,\, \eta=\sqrt{\frac{\sqrt{b+4M}-\sqrt{b-4M}}{\sqrt{b+4M}+\sqrt{b-4M}}}\ .    
\end{align}
One has to emphasise that the expression for the time delay can be expressed in terms of the elliptic integral. However, due to the complexity of the analytical form of the expression, we will not include it in the present article. The simplest analytical form of the deflection angle and time delay of photon can be obtained in the weak field approach. In the next subsection we will show these results.

\subsection{Weak lensing}

In the weak gravitational field approximation, the metric tensor of spacetime can be written as $ g_{\alpha\beta} = \eta_{\alpha\beta} + h_{\alpha\beta} $ and $ g^{\alpha\beta} = \eta^{\alpha\beta} - h^{\alpha\beta} $, where $ \eta_{\mu\nu} $ is the metric tensor in flat space and $ h_{\mu\nu} $ is a small perturbation. Here, $ \eta_{\alpha\beta} = \eta^{\alpha\beta} $, $ h_{\alpha\beta} = h^{\alpha\beta} $, and $ h_{\alpha\beta} h^{\alpha\beta} \to 0 $. According to Ref.~\cite{Kogan10}, the deflection angle of a photon is defined as the difference between the directions of the incoming and outgoing light rays. The deflection angle of the light ray can be expressed as $ \hat{\bf \alpha} = {\bf e}_{\rm out} - {\bf e}_{\rm in} $, where $ {\bf e}_{\rm in} $ and $ {\bf e}_{\rm out} $ are the unit vectors along the spatial component momentum vector $ {\bf p} $ of the "incoming" and "outgoing" photon, respectively. The explicit expression for the deflection angle is given by~\cite{Kogan10}.
\begin{align}\label{alphab}
\hat{\alpha}_b=-\frac{1}{2}\int^{\infty}_{-\infty} \frac{d}{db}\left(h_{tt}+h_{zz}\right)dz\ , 
\end{align}
where $h_{tt}$ and $h_{zz}$ are perturbation in the BBMB spacetime defined as
\begin{align}\nonumber
&h_{tt}=\frac{2M}{r}-\frac{M^2}{r^2}\ ,\\ \label{hab}
&h_{ij}=\left(\frac{2M}{r}+\frac{3M^2}{r^2}\right)\hat{n}_i\hat{n}_j\ , \\\nonumber 
&h_{zz}=\left(\frac{2M}{r}+\frac{3M^2}{r^2}\right) \cos^2\theta\ ,    
\end{align}
where $\hat{n}_i$ is the component of the unit vector with the same direction as the radius vector $r_i = (x,y,z)$ and has the form $\hat n_i = (\cos\phi\sin\theta, \sin\phi\sin\theta, \cos\theta)$ ~\cite{Landau-Lifshitz2}. Using equations~(\ref{alphab}) and (\ref{hab}), the deflection angle of a light ray passing near the BBMB black hole is determined as 
\begin{align}\label{Dangle}
\hat{\alpha}_b=\frac{4M}{b}+\frac{\pi M^2}{4b^2}\ .
\end{align}

Now, we study the observational consequences of gravitational lensing, such as the magnification of image sources, Einstein crosses (rings), and time delays. To do this, we use the lens equation, which relates the angle $\beta$ (the angle between the real position of the source and the lens relative to the observer), the angle $\theta$ (the angle between the apparent image of the source and the observer-lens axis), and the deflection angle $\alpha$:
\[
\beta = \theta - \frac{\Theta_0^2}{\theta},
\]
where $\Theta_0 = \sqrt{\frac{4MD_{\rm ls}}{D_{\rm l} D_{\rm s}}}$ is the Einstein ring radius, with $D_{\rm s}$, $D_{\rm ls}$, and $D_{\rm l}$ being the distances between the observer and the source, the lens and the source, and the observer and the lens, respectively. To consider the image magnification due to lensing, we solve the lens equation:
\[
\theta = \frac{\beta \pm \sqrt{\beta^2 + 4\Theta_0^2}}{2}.
\]
The magnification of the image is given by:
\[
\mu = \left| \frac{\theta}{\beta} \frac{d\theta}{d\beta} \right|,
\]
which can be evaluated for individual images.
\begin{align}\label{mu}
&&\mu_1=\frac{1}{4}\left[\frac{y}{\sqrt{y^2+4}}+\frac{\sqrt{y^2+4}}{y}+2\right]\ ,\\ 
&&\mu_{2}=\frac{1}{4}\left[\frac{y}{\sqrt{y^2+4}}+\frac{\sqrt{y^2+4}}{y}-2\right]\ ,
\end{align}
where $y=\beta/\Theta_0$, the sub indecies ``$1$'' and ``$2$'' denote the primary and secondary images of the source, respectively. The total $\mu=\mu_1+\mu_2$ and the ratio $R=\mu_1/\mu_2$ of magnifications of images are given by 
\begin{align}
\mu=\frac{y^2+2}{y\sqrt{y^2+4}}\ , \qquad R=\left(\frac{\sqrt{y^2+4}+y}{\sqrt{y^2+4}-y}\right)^2\ .
\end{align}
\begin{figure}
\centering\includegraphics[width=\hsize]{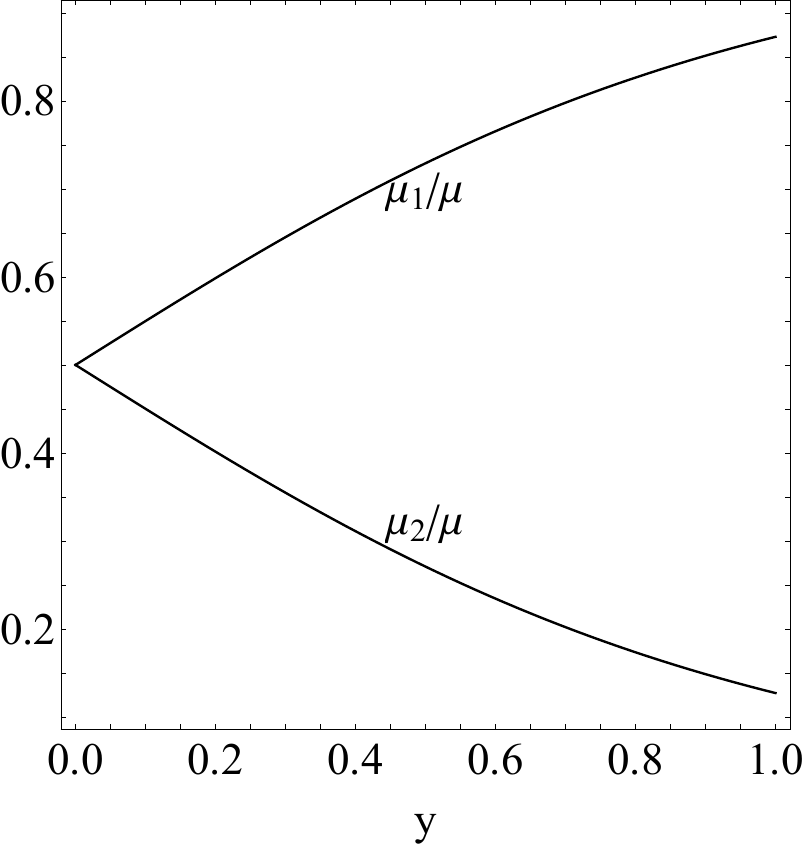}
\caption{The magnification of individual image (left) the total magnification (centre) and the ratio of magnifications (right) as the functions of fractinal angle $y=\beta/\Theta_0$.\label{mag}}
\end{figure}

Figure~\ref{mag} shows the magnification of the primary and secondary images of the source due to weak lensing. This figure illustrates the splitting of the first and second images, or Einstein cross, due to the presence of an external magnetic field relative to the image. In the case of "positive parity," the size of the Einstein cross is larger, while in the case of "negative parity," it is smaller compared to the plasma case.

\begin{figure}
\centerline{\includegraphics[width=\hsize]{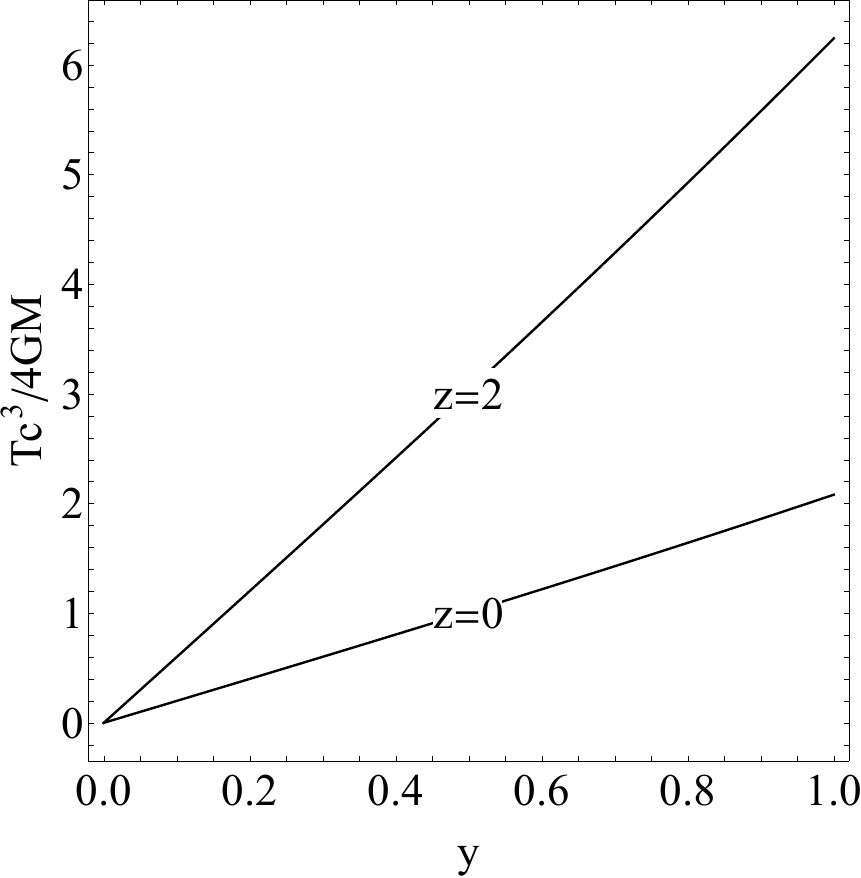}}
\caption{Time delay $T$ of the light-ray as a function of the angle $y=\beta/\Theta_0$ for different values of redshift factor.\label{timedelay}}
\end{figure}

A variable source behind the lensing object produces observable variable images. However, the source and the image will not necessarily vary simultaneously; in general, there will be a time delay between the two events, consisting of two contributions. First, there is a purely geometrical time delay. Second, there is a delay due to the potential of the lensing object, known as the Shapiro time delay. The total time delay, which arises from both the geometry and the gravitational potential, is given by:
\begin{align}
T=\frac{4GM}{c^3}(1+z)\left[\ln\left(\frac{\sqrt{y^2+4}+y}{\sqrt{y^2+4}-y}\right)+\frac{1}{2}y\sqrt{y^2+4}\right]\ .
\end{align}
where $z$ is the redshift of the lensing object. To estimate the value of the time delay for a supermassive black hole with a mass of $10^6 M_\odot$, we can use the following expression:
\[
T \sim 2 \times 10^{-4} \left(\frac{M}{10^6 M_\odot}\right) \text{ s} \ .
\]
The dependence of the time delay due to the gravitational field on the position $y$ for redshift values $z=0$ and $z=2.7$ is illustrated in Fig.~\ref{timedelay}. It is evident that as the angle $\beta$ increases, the time delay of the light ray also increases.

\section{Conclusions\label{Sec:Conclusions}}

In the present paper, we have investigate the novel feature of the BBMB black hole by analysing geodesic motion. First we have considered its thermodynamic properties. It has been shown that unlike the Schwarzschild black hole, the BBMB black hole posses an absolute Hawking temperature of zero, implying it does not emit radiation. However, using the area law, we have shown that the entropy of the BBMB black hole does not become zero. It also shown that since its Hawking temperature is zero, the BBMB black hole does not lose mass over time, meaning its mass remains constant. Consequently, we have summarized that all other thermodynamic quantities of the BBMB black hole which is proportional to the Hawking temperature become zero.

We have examined the motion of both massive and massless particles around the BBMB black hole. By utilizing geodesic motion along constants of motion, we derived analytical expressions for specific energy and specific angular momentum. We identified exact expressions for the characteristic radii around the black hole, as well as the marginally stable and marginally bound circular orbits for massive particles. It was shown that these characteristic radii in the BBMB spacetime are smaller than those predicted in the Schwarzschild spacetime. Additionally, we investigated the energy efficiency of massive particles in the BBMB spacetime and found it can reach up to $8\%$.

We have studied the capture cross section of the massless (photon) and massive particle by the BBMB black hole. From the equation of motion we have found the radial function that plays significant role in finding critical value of the impact parameter of photon and particle. It is found that the critical impact parameter for a photon in the BBMB spacetime is $b = 4M$, and the radius of the photon sphere is $r_{\rm ph} = 2M$. The same results are obtained by analysing the effective potential for photon in the BBMB spacetime. Although, obtained results are compared with those obtained in the Schwarzschild spacetime. These results show that the gravitational properties of the BBMB black hole differ significantly from those of the Schwarzschild black hole. Specifically, the impact parameter for a photon is smaller in the field of a Schwarzschild black hole than in the BBMB case, indicating that gravity around the BBMB black hole is weaker. This is corroborated by the photon sphere's location, being closer in the BBMB spacetime at $2M$ compared to the Schwarzschild spacetime at $3M$. It is shown that capture cross section of the ultra relativistic particle is depends of its energy. 

We have derived the explicit expression for the pericentric precession and the deflection angle of a light by the gravitational object. The explicit expression for the trajectory of massive particle orbiting the BBMB black hole. It is shown that orbit such test particle with elliptic trajectory, and its pericenter always shift in per round which is first predicted in the BBMB spacetime. It is shown that pericentric precession presicted in the BBMB spacetime is slightly less than that predicted in Einstein general theory of relativity. The dependence of the shape of the trajectory of test particle from the eccentricity and expansion parameter has been explicitly analyzed. 

The deflection of the light ray and gravitational lensing effects by the BBMB black hole in strong and in the weak field approximation has been studied. We have provided an analytical computation of the deflection of light and the perihelion precession in the gravitational field of the BBMB black hole spacetime. The derivation of the expression for the deflection angle has been explicitly shown in the first and second order approximations. Using the gravitational lensing equation the magnification of the primary and secondary images has been derived. It is shown that the for small angle $\beta$, it is difficult to distinguish the magnifications of the primary and secondary images, while for largest value of the angle $\beta$ the magnification of the primary image dominates will equal to the total magnification. It is also discussed the time delay of the light ray passing through the BBMB black hole.       

Finally, we have compared all measurable quantities in the BBMB and Schwarzschild spacetime obtained from the geodesic motion. Result are listed in Table \ref{table}. As one can see from the table all quantities are small in the BBMB spacetime than that in the Schwazrschild spacetime due to the weaker gravitational field.

\begin{table}\label{table}
\centering
\caption{Comparison of all measurable quantities in the BBMB and Schwarzschild spacetimes with identical masses.}
\label{parset}
\begin{tabular*}{\columnwidth}{@{\extracolsep{\fill}}llll@{}}
\hline
\multicolumn{1}{@{}l}{Parameters} & BBMB & Schwarzschild\\
\hline
$T$            & $0$ & $(8\pi M)^{-1}$ \\
$S$            & $\pi M^2$ & $4\pi M^2$ \\
$r_{\rm h}$    & $M$ & $2M$ \\
$r_{\rm ms}$   & $4M$ & $6M$ \\
$r_{\rm mb}$   & $(3+\sqrt{5})/2M$ & $4M$\\
$r_{\rm ph}$   & $2M$ & $3M$ \\
$\eta$         & $\sim 6\%$ & $\sim 8\%$ \\
$b$            & $4M$ & $3\sqrt{3}M$ {\qquad\rm for photon}\\
$b$            & $\sqrt{(11+5\sqrt{5})/2}M$ & $4M$ {\,\,\qquad\quad\rm for slow particle}\\
$r_0$          & $(5+\sqrt{5})/12M$   & $4/3M$ {\,\,\quad\quad\rm for slow particle}\\
$\sigma$       & $16\pi M^2$    & $27\pi M^2$ {\qquad\rm for photon} \\
$\sigma$       & $(11+5\sqrt{5})/2\pi M^2$    & $16\pi M^2$ {\qquad\rm for slow particle}\\
$\delta\phi$   & $5\pi (Mm/L)^2$    & $6\pi (Mm/L)^2$ \\
$\alpha$   & $4M/b$    & $4M/b(1+\pi M/16b)$ \\
\hline
\end{tabular*} 
\end{table}


\appendix



\bibliographystyle{elsarticle-harv} 
\bibliography{example}
\end{document}